\documentclass[aps, prl, 10pt, twocolumn, superscriptaddress, preprintnumbers, floatfix]{revtex4-2}

\usepackage[colorlinks=true,breaklinks=true]{hyperref}
\hypersetup{allcolors=[rgb]{0.0 0.0 0.75},linkcolor=[rgb]{0.75 0.05 0.05}}
\usepackage[dvipsnames]{xcolor}
\usepackage{mathtools}
\usepackage{amsmath}
\usepackage{amsfonts}
\usepackage{amssymb}
\usepackage{bbold}
\usepackage{upgreek}
\usepackage{bm}
\usepackage{soul}
\long\def\exclude#1{}
\usepackage{orcidlink}
\usepackage{slashed}
\usepackage{microtype}

\newcommand{\bk}{{\bf k}}

\newcommand{\bB}{{\bf B}}

\newcommand{\br}{{\bf r}}
\newcommand{\bH}{{\bf H}}
\newcommand{\bE}{{\bf E}}
\newcommand{\be}{{\bf e}}

\newcommand{\bb}{{\bf b}}

\newcommand{\bv}{{\bf v}}

\newcommand{\bp}{{\bf p}}
\newcommand{\bj}{{\bf j}}
\newcommand{\bU}{{\bf U}}

\begin{document}
\preprint{TIFR/TH/26-31}
\title{Phase-Twist Instability in a Coprecessing Neutrino Plasma}

\author{Damiano F.\ G.\ Fiorillo \orcidlink{0000-0003-4927-9850}}
\email{damiano.fiorillo@gssi.it}
\affiliation{Gran Sasso Science Institute (GSSI), Viale Francesco Crispi 7, 67100 L’Aquila, Italy}
\affiliation{Istituto Nazionale di Fisica Nucleare (INFN), Sezione di Napoli, Complesso Universitario di Monte Sant'Angelo, Via Cintia, 80126 Napoli, Italy}

\author{Dwaipayan Mukherjee \orcidlink{0009-0005-2204-2249}} 
\affiliation{Tata Institute of Fundamental Research, Homi Bhabha Road, Mumbai, 400005, India}

\author{Basudeb Dasgupta \orcidlink{0000-0001-6714-0014}} 
\affiliation{Tata Institute of Fundamental Research, Homi Bhabha Road, Mumbai, 400005, India}

\author{Georg~G.~Raffelt~\orcidlink{0000-0002-0199-9560}}
\affiliation{Max-Planck-Institut f\"ur Physik, Boltzmannstra\ss e~8, 85748 Garching, Germany}

\begin{abstract}
Dense neutrino plasmas support collective flavor motion. We derive the first explicit dispersion relation about a flavor-nondiagonal background, focusing on the coprecession solution underlying spectral-split formation during the prompt neutrino burst of an electron-capture supernova. We show that slow phase modulations are unstable Goldstone modes, causing the coprecession plane to wobble with increasing amplitude. The instability emerges in the hydrodynamic limit, governed by conservation laws and without resolving flavor-precession resonances. Coprecessing neutrino plasmas resemble magnetized plasmas: a coherent flavor field drives flavor isospins to precess, and the phase-twist instability is the flavor analog of the firehose instability.
\end{abstract}

\maketitle

\textbf{\textit{Introduction}}---In the dense outflows of supernovae and neutron-star mergers, neutrinos undergo collective flavor evolution, driven by the coherent weak field they themselves generate~\cite{Pantaleone:1992eq, Samuel:1993uw, Pantaleone:1994ns, Samuel:1995ri, Duan:2006an}. Several reviews cover both theoretical foundations and astrophysical applications~\cite{Duan:2009cd, Duan:2010bg, Tamborra:2020cul, Volpe:2023met, Johns:2025mlm, Raffelt:2025wty, Fiorillo:2026gzy}. This collisionless yet correlated neutrino state behaves as a plasma in flavor space~\cite{Fiorillo:2025zio}, and the focus has been on flavor-diagonal plasmas, which still support waves of flavor coherence~\cite{Izaguirre:2016gsx, Capozzi:2017gqd, Yi:2019hrp, Fiorillo:2024bzm, Fiorillo:2024pns, Fiorillo:2025npi, Fiorillo:2025zio, Johns:2025yxa, Kost:2026ckc}. These modes can be unstable, driving strong flavor motion on small scales.

\begin{figure}[t!]
    \centering
    \includegraphics[width=\columnwidth]{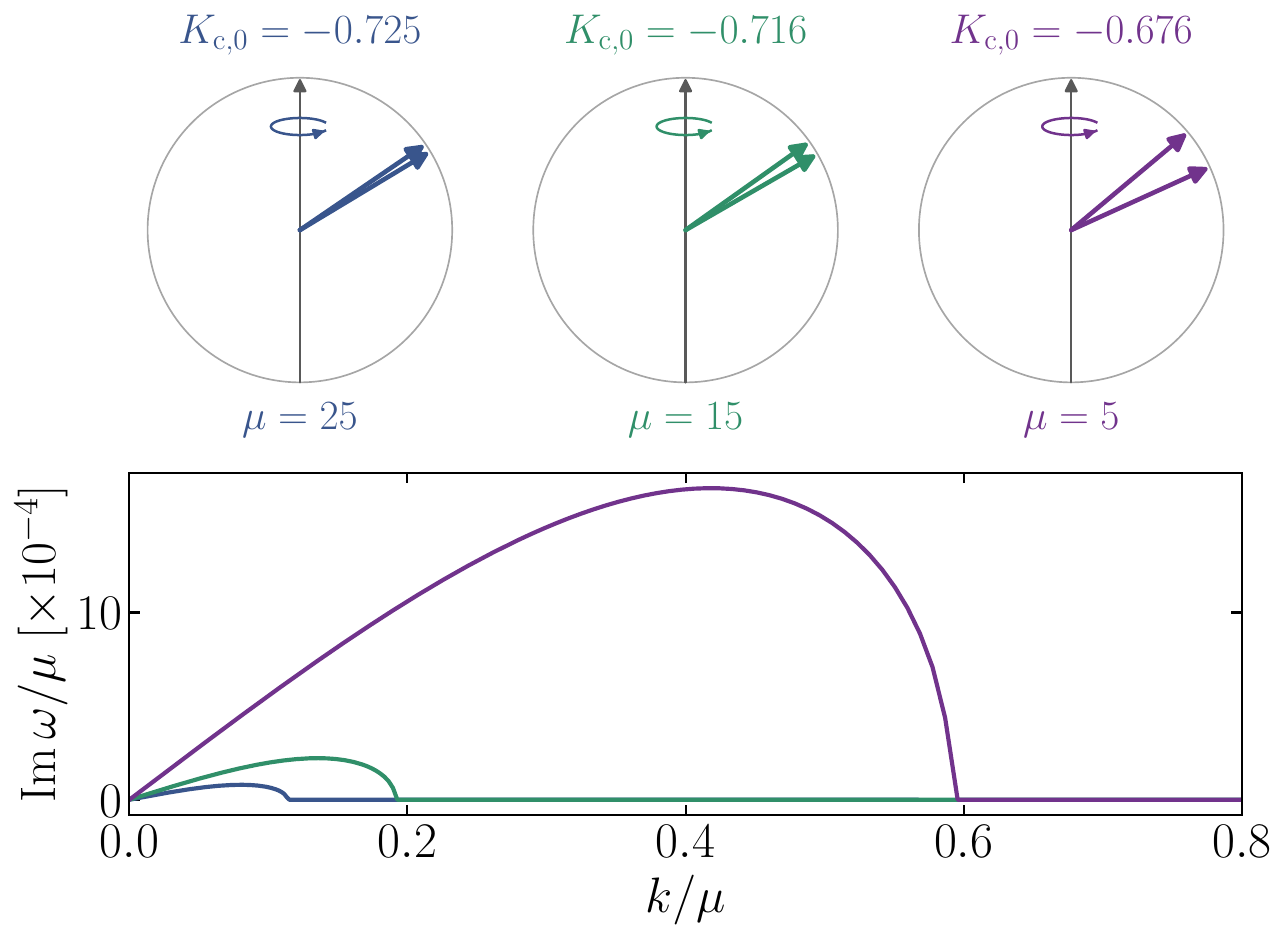}
    \caption{Flavor coprecession in space with wavevector $K_{\rm c,0}$. Our minimal setup uses two neutrino modes $\vec{P}_a$ and~$\vec{P}_b$, each of length $1$, vacuum oscillation frequencies $w_a=1$ and $w_b=0.4$, and momentum orientations relative to the axial symmetry axis of $v_a=1$ and $v_b=0.9$. Neutrino--neutrino interaction strength $\mu$ is varied; see End Matter for the detailed construction.  
    \textit{Top:} From left to right, decreasing $\mu$. \textit{Bottom:} Growth rate of the unstable mode, which is linear in $k$ in the hydrodynamic regime and vanishes at large wavenumbers.}
    \label{fig:scheme}
\end{figure}

We turn instead to the stability of perturbations about a flavor-nondiagonal plasma, which already possesses a coherent flavor field. We focus on a coprecession solution \cite{Raffelt:2007xt, Duan:2008fd}, in which the flavor polarization vectors span a common plane that rotates around the mass direction (Fig.~\ref{fig:scheme}). For coprecessions in the fast limit of vanishing neutrino masses, the rotation axis is instead fixed by initial conditions~\cite{Duan:2021woc, Liu:2025muc, Fiorillo:2026jgw}. In a one-dimensional system, the corotation phase varies as $\Phi=\Omega_{\rm c, 0} t-K_{\rm c,0} z$. 

We further restrict to coprecessions with all polarization vectors on one side of the mass axis (Fig.~\ref{fig:scheme}). Such configurations underlie spectral-split formation during the prompt $\nu_e$ burst of electron-capture supernovae \cite{Duan:2007sh, Dasgupta:2008cd}. Our companion paper~\cite{Mukherjee:2026xyw} explores how the instability appears in this process, motivating our theoretical analysis. In a  different context, disturbances about generic flavor-nondiagonal backgrounds were proposed in Refs.~\cite{Johns:2025yxa, Kost:2026ckc}, and instabilities of phase-randomized nondiagonal ensembles were studied in Ref.~\cite{Majumder:2026thb}. We here provide the first explicit dispersion relation around a consistent nondiagonal background.

A key new result of our analysis is that 
these one-sided states maximize the corotating flavor energy and are thus stable against homogeneous perturbations which conserve it. Uniform phase rotations cost no energy, making slow phase twists natural disturbances to consider: they need not be stable because inhomogeneous perturbations do not conserve flavor energy \cite{Fiorillo:2024fnl}. When these twists vary slowly compared with microscopic flavor precession, the polarization vectors adjust locally to slightly shifted coprecession states. 

In this hydrodynamic limit, the evolution is described by the macroscopic phase $\Phi$ and governed by lepton-number conservation. We consider a perturbation
\begin{equation}
    \delta \Phi=\delta \Phi_0 e^{-i\omega t+i k z},
\end{equation}
describing a wobble of the rotation plane. This phase-twist is a Goldstone mode from rotational symmetry around the mass axis: we prove its generic instability with a growth rate $\gamma=\mathrm{Im}(\omega)\propto k$ (Fig.~\ref{fig:scheme} bottom).

At larger $k$, perturbations resolve individual flavor precession, taking us into the kinetic regime. We therefore derive the flavor susceptibility of the coprecessing plasma, generalizing the flavor-diagonal case \cite{Fiorillo:2024qbl,Fiorillo:2024bzm, Fiorillo:2024uki, Fiorillo:2024pns, Fiorillo:2025ank, Fiorillo:2025npi, Fiorillo:2025zio, Fiorillo:2026byh, Fiorillo:2026tee}. Cherenkov resonances give way to cyclotron resonances when the wave frequency matches the neutrino flavor-precession frequency. Such resonant spin response was identified in the early literature~\cite{Raffelt:2008hr,Wu:2011yi}, but only as a mechanical resonance of homogeneous spin systems; here we deal instead with the kinetic response of a flavor-nondiagonal plasma, where resonances arise through the Doppler-shifted frequency along each neutrino trajectory. The instability ultimately disappears at sufficiently large $k$
(Fig.~\ref{fig:scheme} bottom).

The overall picture is surprisingly similar to the firehose instability of a magnetized plasma~\cite{Parker:1958zzb}, where magnetic field lines buckle under an anisotropic pressure in the hydrodynamic regime, while kinetic effects stabilize them at short wavelengths.

{\bf\textit{Evolution and conservation laws}}---To set~the~stage, we restrict to a two-flavor system. A neutrino mode with momentum $\bp$ is described by a $2{\times}2$ density matrix $\rho_\bp=(n_\bp+\vec{P}_\bp\cdot \vec{\sigma})/2$, where $n_\bp$ is the total occupation number, \smash{$\vec{P}_\bp$} the flavor polarization vector, and $\vec{\sigma}$ the Pauli matrices. We neglect antineutrinos, as appropriate for the prompt $\nu_e$ burst. We denote flavor-space vectors by arrows and phase-space vectors in boldface. $\vec B$ is the unit vector along the mass direction and $P_{\parallel,\bp}=\vec B\cdot\vec P_\bp$ the projection of \smash{$\vec{P}_\bp$} onto this direction.

For a local stability analysis, we assume a homogeneous background and restrict to one-dimensional, axially symmetric evolution along $z$, with angular dependence only through $v=\cos\theta$. We retain $\bp$ as shorthand for $v$ and energy, so that the evolution of $\vec P_\bp(t,z)$ is determined by the mean-field kinetic equation~\cite{Dolgov:1980cq, rudzsky1990kinetic, Sigl:1993ctk,Vlasenko:2013fja,Serreau:2014cfa,Fiorillo:2024fnl,Fiorillo:2024wej,Fiorillo:2026byi}
\begin{equation}\label{eq:general_evolution}
    (\partial_t+v\partial_z)\vec{P}_\bp=[w_\bp\vec{B}+\mu(\vec{P}_0-v\vec{P}_1)]\times \vec{P}_\bp,
\end{equation}
where \smash{$w_\bp=\delta m^2/2E_\bp$} is the vacuum energy splitting, acting as an external field along \smash{$\vec{B}$}, and $\vec{P}_n=\sum_\bp \vec{P}_\bp v^n$ are the angular moments. The sum is understood as a phase-space integral \smash{$n_\nu^{-1}\int d^3\bp/(2\pi)^3$} in the continuous case. We define the neutrino--neutrino interaction strength as $\mu=\sqrt{2}G_{\rm F} n_\nu$, with $\vec P_n$ dimensionless and of order unity. Only the combination $\mu(\vec P_0-v\vec P_1)$ is physical. Henceforth, we absorb $\mu$ into the units of space and time.

The kinetic equation immediately implies the continuity equation
\begin{equation}\label{eq:lepton_number_conservation}
    \partial_t L_0+\partial_z L_1=0,
\end{equation}
where $L_n=P_{\parallel,n}=\sum_\bp v^n P_{\parallel,\bp}$, so that the total mass-state lepton number is conserved. In our $(t,z)$ description, the first two moments are precisely the temporal and spatial components of the lepton-number four-current restricted to $1+1$ dimensions: $L^\mu=(L_0,L_1)$.
In contrast to mass-state lepton number, the energy functional
\begin{equation}
    \mathcal{E}=\sum_\bp w_\bp P_{\parallel,\bp}+\frac{|\vec{P}_0|^2-|\vec{P}_1|^2}{2}
\end{equation}
does not obey a conservation law. In a nonuniform system
\begin{equation}
    \partial_t\mathcal{E}+\partial_z\biggl(\sum_\bp w_\bp vP_{\parallel,\bp}\biggr)=\vec{P}_1\cdot\partial_z\vec{P}_2-\vec{P}_0\cdot \partial_z\vec{P}_1.
\end{equation}
Refractive energy is not separately conserved because of energy exchange with the kinetic neutrino motion \cite{Fiorillo:2024fnl}, a central ingredient of the new instability.

{\bf\textit{Flavor coprecession}}---Several classes of exact solutions of Eq.~\eqref{eq:general_evolution} have been found~\cite{Hannestad:2006nj, Raffelt:2007cb, Raffelt:2007xt, Raffelt:2011yb,Dasgupta:2017oko,Johns:2019izj,Padilla-Gay:2021haz,Fiorillo:2023mze, Fiorillo:2023hlk, Liu:2025muc, Fiorillo:2026jgw, Fiorillo:2026att, Fiorillo:2026lyz}. The simplest nontrivial case is a self-consistent coplanar configuration first identified in the theory of spectral-split formation \cite{Raffelt:2007cb, Raffelt:2007xt,Duan:2008fd}.  All \smash{$\vec{P}_\bp$} remain coplanar, with their common plane rotating around \smash{$\vec B$}, its phase advancing linearly in space and time.

Henceforth we will work in the corotating frame, using $\vec{P}_\bp$ for the polarization vector in this frame and 
$\vec{P}_\bp^{\,\rm lab}$ for the original vector. We may then write the coprecession solution in the form
\begin{equation}
  \vec{P}_\bp^{\,\rm lab}=R\bigl[\Phi(t,z)\bigr]\vec{P}_\bp,
\end{equation}
where $R[\Phi]$ denotes a rotation around \smash{$\vec{B}$} by an angle $\Phi$. 

With $\partial_t \Phi(t,z)=\Omega_{\rm c}(t,z)$ and $\partial_z \Phi(t,z)=-K_{\rm c}(t,z)$, the equation for the corotating \smash{$\vec{P}_\bp(t,z)$} is
\begin{equation}
    (\partial_t+v\partial_z)\vec{P}_\bp=\left[\big(w_\bp-\Omega_{\rm c}+v K_{\rm c}\big)\vec{B}+\vec{P}_0-v\vec{P}_1\right]\times \vec{P}_\bp.
\end{equation}
For an exact coprecession, $\Omega_{\rm c}(t,z)=\Omega_{\rm c,0}$ and $K_{\rm c}(t,z)=K_{\rm c,0}$ are constant and every \smash{$\vec{P}_\bp$} is stationary and homogeneous. This solution requires the polarization vectors to be collinear with their Hamiltonian vectors, \smash{$\vec{P}_\bp=\alpha_\bp \vec{H}_\bp$},
\begin{equation}
    \vec{H}_\bp=(w_\bp-\Omega_{\rm c,0}+v K_{\rm c,0})\vec{B}+\vec{P}_0-v\vec{P}_1.
\end{equation}
A consistency condition immediately follows from the requirement $\sum_\bp \alpha_\bp \vec{H}_\bp v^n=\vec{P}_n$ with $n=0$ or~1. The moments \smash{$\vec{P}_n$}, and in particular their components transverse to \smash{$\vec{B}$}, are the classical flavor fields introduced in the linear response theory of Refs.~\cite{Fiorillo:2024bzm, Fiorillo:2024uki, Fiorillo:2025npi}. The consistency condition is then akin to a field equation, relating the overall flavor field to its microscopic sources $\vec{P}_\bp$.

One notable property of coprecessions is that they extremize the corotating flavor energy $\mathcal{H}[\vec{P}_\bp; \Omega_{\rm c},K_{\rm c}]=\mathcal{E}-\Omega_{\rm c}L_0+K_{\rm c} L_1$. One immediately verifies that \smash{$\delta \mathcal{H}/\delta \vec{P}_\bp=\vec{H}_\bp$}. Since \smash{$\vec{P}_\bp=\alpha_\bp \vec{H}_\bp$}, any length-preserving variation satisfies
\smash{$\delta\vec{P}_\bp\cdot\vec{H}_\bp=0$}. Therefore, the coprecession is an extremum of \smash{$\mathcal{H}_0[\vec{P}_\bp]=\mathcal{H}[\vec{P}_\bp;\Omega_{\rm c,0},K_{\rm c,0}]$}. 

In the End Matter, we prove the even stronger result that, when all $\alpha_\bp>0$, the coprecession is a maximum of $\mathcal{H}_0$. This applies to the prompt $\nu_e$ burst, where all neutrinos start out aligned without a crossing. The only deformation of the solution that costs exactly zero energy is a change in the common phase $\Phi$, namely \smash{$\delta\vec{P}_\bp=\delta\Phi \vec{B}\times \vec{P}_\bp$}. Any other perturbation lowers \smash{$\mathcal{H}_0[\vec{P}_\bp]$}.

The maximum property immediately shows that any homogeneous perturbation, \smash{$\partial_z\delta\vec{P}_\bp=0$}, must be stable. Indeed, in a homogeneous system $\partial_t \mathcal{H}_0=0$, so energy conservation prevents unstable growth. Inhomogeneities, however, generally change the neutrino--neutrino interaction energy. Although of small quantitative relevance for flavor-diagonal backgrounds~\cite{Fiorillo:2024fnl}, this effect is crucial here, enabling unstable inhomogeneous modes.

A final property of coprecessions is crucial to their instability. Consider a coprecession $(\Omega_{\rm c,0},K_{\rm c,0})$ and an infinitesimally displaced 
one $(\Omega_{\mathrm{c},0}+d\Omega_{\rm c},K_{\mathrm{c},0}+dK_{\rm c})$, for which the lepton-number moments change by $d L_0$ and $d L_1$. The new coprecession extremizes $\mathcal{H}_0-d \Omega_{\rm c}L_0+d K_{\rm c} L_1$. At the neighboring coprecession, stationarity therefore implies
\begin{equation}
    d \mathcal{H}_0=d \Omega_{\rm c} d L_0-d K_{\rm c} d L_1.
\end{equation}
Since $d L_0$ and $dL_1$ 
are linear in $d\Omega_{\rm c}$ and $dK_{\rm c}$, integrating this relation to quadratic order for a small but nonzero displacement $(\Delta\Omega_{\rm c},\Delta K_{\rm c})$ gives
\begin{equation}\label{eq:change_in_energy_finite}
    2\Delta \mathcal{H}_0=\Delta \Omega_{\rm c} \Delta L_0-\Delta K_{\rm c} \Delta L_1.
\end{equation}
We verify this relation explicitly in the End Matter.

These relations describe conjugate Hamiltonian variables. Define $F(\Omega_{\rm c},K_{\rm c})=\mathcal{H}[\vec{P}_\bp(\Omega_{\rm c},K_{\rm c});\Omega_{\rm c},K_{\rm c}]$, where $\vec{P}_\bp(\Omega_{\rm c},K_{\rm c})$ is the corresponding coprecession solution. By stationarity,
\begin{equation}\label{eq:first_law_thermodynamics}
    dF=-L_0 d\Omega_{\rm c}+L_1 dK_{\rm c},
\end{equation}
so $F$ acts as a generating potential. 
There is also a point of contact with the neutrino thermodynamics of Ref.~\cite{Johns:2023jjt}, where flavor evolution is postulated to pass through local mixing equilibria, with flavor coprecession providing one example. However, Eq.~\eqref{eq:first_law_thermodynamics} is not thermodynamical, and is proven from first principles.

{\bf\textit{Hydrodynamic instability}}---It is now immediate to show that an instability for a slowly varying perturbation is certain to exist. We start from a perturbation of zero energy cost, corresponding to an infinitely slow spacetime modulation of the coprecession plane. This is a Goldstone mode, deriving from the symmetry of the equations around the mass axis $\vec B$. The necessity of such a mode was also noted in Ref.~\cite{Johns:2025yxa}, but its dispersion and, crucially, instability was not discussed. We thus expect a branch of perturbations around the coprecession that becomes static in the infinite-wavelength limit. What becomes of this Goldstone mode for slow variations?

If the phase $\Phi(t,z)=\Omega_{\rm c,0} t-K_{\rm c,0} z+\delta\Phi(t,z)$ contains the small, slowly varying modulation $\delta \Phi=\delta \Phi_0 e^{-i\omega t +i k z}$,
then $\Omega_{\rm c}=\partial_t\Phi$ and $K_{\rm c}=-\partial_z\Phi$ become slowly varying, and so do $L_0$ and $L_1$. The induced changes are, to linear order, $\delta L_0(t,z)=\partial_{\Omega_{\rm c}} L_0 \partial_t \delta\Phi - \partial_{K_{\rm c}}L_0 \partial_z\delta\Phi$ and similarly for $\delta L_1$. Substitution into the continuity Eq.~\eqref{eq:lepton_number_conservation} gives
\begin{equation}\label{eq:hydro_evolution}
  \partial_{\Omega_{\rm c}} L_0\,\partial_t^2\delta\Phi
-\partial_{K_{\rm c}} L_1\,\partial_z^2\delta\Phi
+\left(\partial_{\Omega_{\rm c}} L_1-\partial_{K_{\rm c}} L_0\right)
\partial_t\partial_z\delta\Phi=0.
\end{equation}
For a plane-wave modulation  $\delta\Phi\propto e^{-i\omega t+ikz}$, we obtain a quadratic dispersion relation 
\begin{equation}\label{eq:HydroDispersion}
    \omega^2 \Upsilon_{00}+k^2\Upsilon_{11}+\omega k(\Upsilon_{01}+\Upsilon_{10})=0.
\end{equation}
The coefficients $\Upsilon_{00}=\partial_{\Omega_{\rm c}} L_0$, $\Upsilon_{10}=\partial_{K_{\rm c}} L_0$, $\Upsilon_{01}=-\partial_{\Omega_{\rm c}} L_1$, and $\Upsilon_{11}=-\partial_{K_{\rm c}} L_1$ depend only on the background family of coprecessions and can therefore be obtained without solving the perturbation equations.

The two components of this conserved current are the density $L_0$ and the $z$-flux $L_1$, while $\Omega_{\rm c}$ and $K_{\rm c}$ are the corresponding time and space gradients of $\Phi$. Generalization beyond axisymmetry obtains directly from Lorentz invariance: we promote the lepton-number flux to a four-vector \smash{$L^\mu=(L_0,\boldsymbol{L})$} with $\boldsymbol{L}=\sum_\bp P_{\parallel,\bp}\bv$, where $\bv=\bp/|\bp|$, and introduce $\Lambda^\mu=(\Omega_{\rm c},\boldsymbol{K}_{\rm c})$ with $\boldsymbol{K}_{\rm c}=-\boldsymbol{\nabla}\Phi$. If we call as usual $K^\mu=(\omega,\bk)$ for the perturbation, the dispersion relation becomes $\Upsilon_{\mu\nu} K^\mu K^\nu=0$ with
\begin{equation}
    \Upsilon_{\mu\nu}=\frac{\partial L_\nu}{\partial \Lambda^\mu}=-\frac{\partial^2 F}{\partial \Lambda^\mu \partial \Lambda^\nu}.
\end{equation}
This generalizes the coefficients stated after Eq.~\eqref{eq:HydroDispersion}.

The two solutions of the quadratic dispersion relation Eq.~\eqref{eq:HydroDispersion} can be written as $\omega=u_\pm k$, where $u_\pm$ are either both real or complex conjugates, the latter implying an instability. It is the same maximum property that enforced stability at $k=0$ that now enforces instability at nonzero $k$. Suppose that $u_+$ is real, so there is a small real perturbation $\Delta L_0$ and $\Delta L_1$. Then the conservation Eq.~\eqref{eq:lepton_number_conservation} implies that the perturbed lepton numbers satisfy $\Delta L_1=u_+ \Delta  L_0$. At the same time, we also have $\Delta\Omega_{\rm c}=u_+\Delta K_{\rm c}$ by definition. Therefore, from Eq.~\eqref{eq:change_in_energy_finite}, the energy change associated with the perturbation must vanish, $\Delta \mathcal{H}_0=0$. But for nonzero $k$, no perturbation costs zero energy: the pure phase variation is always accompanied by an in-plane distortion, and that always lowers the energy. Hence, $u_\pm$ must be complex. In other words, the evolution Eq.~\eqref{eq:hydro_evolution} for $\delta \Phi$ is elliptic, rather than hyperbolic, leading to growth.

In fact, our description borrows heavily from that of density waves in superfluid hydrodynamics, which are also described by a single variable, the phase $\Phi$ of the condensate wavefunction. The hydrodynamic dispersion relation Eq.~\eqref{eq:HydroDispersion} descends from an effective action expanded in the slow gradients
\begin{equation}\label{eq:Seff}
    S_{\mathrm{eff}}[\delta\Phi]=\frac{1}{2}\int dt dz \Upsilon^{\mu\nu}\partial_\mu\delta\Phi\, \partial_\nu\delta\Phi,
\end{equation}
so the tensor $\Upsilon^{\mu\nu}$ plays the role of the superfluid stiffness. 

There are two limits in which the instability disappears. The first is when all modes have the same velocity $v$. Stability follows directly from Eq.~\eqref{eq:general_evolution}, since the convective derivative $\partial_t+v\partial_z$ then becomes a total derivative along the trajectory, reducing the problem to homogeneous evolution, where energy conservation forbids instability. Therefore, for a narrow spread $\Delta v$, the growth rate must be proportional to $\Delta v$.

The second stable limit is reached when all polarization vectors become aligned. This occurs either in the fast limit ($w_\bp=0$) with alignment along any direction, or generically for $w_\bp\neq 0$ along \smash{$\vec{B}$}. For $\alpha_\bp>0$, these systems have no crossing, precluding instability~\cite{Dasgupta:2021gfs} because they maximize total lepton number~\cite{Johns:2024esf, Fiorillo:2024bzm}. For a coprecession whose polarization vectors have a small angular spread $\Delta\theta\ll1$, the growth rate must be proportional to $\Delta\theta$. Together with the $\Delta v$ scaling, this implies $\mathrm{Im}(u_+)\sim\Delta v\Delta\theta$, where $\gamma=\mathrm{Im}(\omega)=\mathrm{Im}(u_+)k$.

{\bf\textit{Beyond hydrodynamics}}---The linear increase of $\gamma$ with $k$ ends as the hydrodynamic approximation breaks down. This occurs when the perturbation begins to resolve microscopic dynamical scales, here set by the precession frequency $|\vec{H}_\bp|$ of a typical $\vec{P}_\bp$. Equivalently, the frequency of the perturbation felt by an individual neutrino along its trajectory, $\omega'=\omega-kv\simeq k(u_\pm-v)$, becomes comparable with \smash{$|\vec{H}_\bp|$}. Corrections to hydrodynamic results are then controlled by $k \Delta v/|\vec{H}_\bp|$. For $\Delta v\to 0$, the dispersion remains exactly linear in $k$ since all modes are simply convected with $\omega'=\omega-kv$.

The linear dispersion relation $\omega=u_\pm k$ thus receives higher-order corrections in $k$. To constrain their structure, we note the invariance of the dispersion relation under $(\omega,k)\to (-\omega,-k)$. This follows from the invariance of the equations and coprecessing background under $(t,z)\to( -t,-z)$ combined with reflection of the flavor polarization vectors about the coprecession plane.

With $u_\pm=u_R\pm iu_I$, the linear form $\omega=u_\pm k$ can be expressed as
$(\omega-u_{R}k)^2=-u_I^2 k^2$. Assuming $|u_I|\ll |u_R|$, the leading higher-order correction may be written as
\begin{equation}\label{eq:dispersion_relation_hydro_expansion}
    (\omega-u_R k)^2=-u_I^2 k^2+\beta k^4+\ldots
\end{equation}
since terms of order $k^{4-n}(\omega-u_R k)^{n}$ with $n\geq 1$ are parametrically smaller. Stabilization requires $\beta>0$, as confirmed by our two-beam solution in Fig.~\ref{fig:scheme}. Our earlier arguments imply $\beta\sim \Delta v^4/|\vec{H}_\bp|^2$. Therefore, the increase of $\gamma$ turns over at a typical \smash{$k\sim |\vec{H}_\bp|\Delta \theta/\Delta v$}, with a maximum growth rate \smash{$\gamma_{\rm max}\sim |\vec{H}_\bp|\Delta \theta^2$}. 

This structure is realized explicitly by the two-mode solution in Fig.~\ref{fig:scheme}. For small $\Delta \theta$, the instability disappears \textit{before} reaching the purely kinetic regime $k\sim |\vec{H}_\bp|/\Delta v$ and it suffices to consider the first non-hydrodynamic correction to describe the entire unstable range. 
One direct prediction of our dimensional estimates is that even in the nearly-aligned regime, $w_\bp\ll \mu$, 
where $\Delta\theta\sim w_\bp/\mu$, the instability persists with a parametrically small growth rate $\gamma\sim w_\bp^2/\mu$. The unstable mode disappears by merging with its complex conjugate at a degeneracy point, whose analytic structure is discussed in Refs.~\cite{Yi:2019hrp,Fiorillo:2024uki,Fiorillo:2026zcm}.

{\bf\textit{Kinetic regime}}---To go beyond the parametric estimate for $\beta$ and describe the instability fully, we linearize Eq.~\eqref{eq:general_evolution}. A small perturbation $\delta\vec{P}_\bp$ about a coprecession obeys, with \smash{$\delta\vec{H}_\bp=\delta\vec{P}_0-v\delta\vec{P}_1$},
\begin{equation}\label{eq:perturbation_evolution}
    (\partial_t+v\partial_z)\delta\vec{P}_\bp=\vec{H}_\bp\times\delta\vec{P}_\bp-\alpha_\bp \vec{H}_\bp\times \delta\vec{H}_\bp.
\end{equation}
The phase perturbation $\delta \Phi$ considered in the hydrodynamic regime is a special form of perturbation. In Eq.~\eqref{eq:perturbation_evolution}, the first term describes precession of individual neutrino modes around the background field $\vec{H}_\bp$, analogous to cyclotron motion in a magnetized plasma. The second term accounts for the collective field $\delta\vec{H}_\bp$ sourced by the neutrino perturbations. Thus, even with \smash{$\delta\vec{H}_\bp$} neglected, the microscopic flavor dynamics is precession around $\vec H_\bp$ rather than free streaming. It is this microscopic precession, absent for flavor-diagonal backgrounds, that drives the analogy with a magnetized plasma.

This analogy shows up directly in the flavor susceptibility and dispersion relation. Since $\delta\vec{P}_\bp\perp\vec{H}_\bp$ by length conservation, imposing \smash{$\delta\vec{P}_\bp\propto e^{-i\omega t+ i k z}$} in Eq.~\eqref{eq:perturbation_evolution} and taking the cross product with \smash{$\vec{H}_\bp$} gives
\begin{equation}\label{eq:explicit_solution}
    \delta\vec{P}_\bp=\frac{i\alpha_\bp (\omega-kv)\vec{H}_\bp \times \delta\vec{H}_\bp+\alpha_\bp\vec{H}_\bp\times (\delta\vec{H}_\bp\times\vec{H}_\bp)}{|\vec{H}_\bp|^2-(\omega-kv)^2}.
\end{equation}
Summing over $\bp$, we can relate the total perturbation from the particles on the left-hand side to the field perturbation $(\delta \vec{P}_0,\delta \vec{P}_1)$ on the right-hand side; the two need of course be equal. The resulting flavor susceptibility is reported in full in the End Matter.

Beyond the explicit form of the susceptibility implied by these expressions, the crucial feature is that its resonances are shifted from $\omega-kv=0$ in a flavor-diagonal plasma to $\omega-kv=\pm |\vec{H}_\bp|$. A resonant wave-particle interaction therefore occurs when the perturbation frequency, in the frame comoving with the particle, matches the flavor precession frequency. This is precisely the cyclotron-resonance condition for a magnetized plasma, where $|\omega-kv|$ matches the cyclotron frequency or one of its harmonics. In our SM~\cite{Supplemental-Material}, we review the basic theory of magnetized plasmas to establish this analogy more explicitly.

{\bf\textit{Discussion}}---We have derived the first explicit dispersion relation for perturbations about a flavor-nondiagonal background, focusing on one-sided coprecessions. Flavor-diagonal backgrounds resemble unmagnetized plasmas, whereas the more general cases are akin to magnetized plasmas, possessing a coherent flavor field that can become unstable to small, self-consistent perturbations.

We focus on the hydrodynamic limit of slow phase variations, meaning long-wavelength, low-frequency collective dynamics in which rapid microscopic precession is averaged over. Here, ``hydrodynamic'' does not imply collisional relaxation or local thermal equilibrium. Similarly, fluid descriptions of collisionless magnetized plasmas can emerge by averaging over rapid gyromotion.

Neutrinos moving in different directions sample the modulation differently and acquire relative twists in the coprecession plane. Our energy argument proves generic instability for one-sided configurations (all polarization vectors on the same side of the mass axis). When this condition fails ($\alpha_\bp$ have different signs), other instabilities may arise. For example, when $\mu\gg w_\bp$, the vectors become collinear but exhibit a spectral crossing, allowing instability already in the flavor-diagonal limit.

The wobbling of the coprecession plane is directly analogous to the firehose instability of a magnetized plasma (see Supplemental Material, SM~\cite{Supplemental-Material}). There, magnetic field lines buckle in response to a large pressure parallel to the field, whose centrifugal stress amplifies the deformation.

The analogy persists all the way to kinetic scales, where perturbations begin to resolve individual flavor precession. The plasma response develops wave-particle resonances when the wave frequency matches a neutrino's flavor-precession frequency, replacing the Cherenkov resonances of flavor-diagonal plasmas~\cite{Fiorillo:2024bzm,Fiorillo:2025npi}.

Our proof that the Goldstone mode is generically unstable does not exclude additional instabilities in the purely kinetic regime. Identifying these is much harder and requires the generalized flavor susceptibility given in the End Matter. Our main result is that a hydrodynamic instability always exists for $\alpha_\bp>0$. 

This finding complements existing criteria for flavor-diagonal backgrounds: the necessity of a spectral crossing~\cite{Dasgupta:2021gfs}, later understood as a constraint from lepton number conservation~\cite{Johns:2024esf,Fiorillo:2024bzm}; Morinaga's sufficiency theorem for fast instabilities~\cite{Morinaga:2021vmc} and its physically transparent proof~\cite{Fiorillo:2024bzm}; the Nyquist criteria for homogeneous instabilities~\cite{Fiorillo:2023hlk,Fiorillo:2024dik}; sufficient and necessary conditions for fast, slow, and mixed instabilities~\cite{Dasgupta:2025quc}; the criterion for collisional instabilities~\cite{Johns:2021qby,Fiorillo:2025zio,Fiorillo:2026gzy}; and the kinematic criterion for slow instabilities~\cite{Fiorillo:2025zio,Fiorillo:2025kko}.

Phenomenologically, we determine the characteristic spatial and temporal scales of the instability in terms of the angular spread of flavor isospins, neutrino density, and directional spread of neutrinos. These scalings are confirmed by the exact two-mode solution in Fig.~\ref{fig:scheme}.

The phase-twist instability of a flavor-nondiagonal background and the explicit dispersion relation provide a concrete realization of more general flavor waves envisioned in Refs.~\cite{Johns:2025yxa, Kost:2026ckc}. As our own prime motivation, we show in a companion paper~\cite{Mukherjee:2026xyw} that the prompt $\nu_e$ burst of an electron-capture supernova develops the phase-twist instability before spectral splits can form.

{\bf\textit{Acknowledgments}}---The work of DFGF is funded by the European Union (ERC, NEPTUNE, No.\ 101301124). Views and opinions expressed are those of the author(s) only and do not necessarily reflect those of the European Union or the European Research Council. Neither the European Union nor the granting authority can be held responsible for them. DFGF acknowledges partial support by Istituto Nazionale di Fisica Nucleare (INFN) through the Theoretical Astroparticle Physics (TAsP) project. GGR acknowledges partial support by the German Research Foundation (DFG) through the Collaborative Research Centre ``Neutrinos and Dark Matter in Astro- and Particle Physics (NDM),'' Grant SFB--1258--283604770 and under Germany’s Excellence Strategy through the Cluster of Excellence ORIGINS EXC--2094--390783311. DM and BD acknowledge support from the Department of Atomic Energy, Government of India, under Project Identification Number RTI-4012.

\bibliographystyle{bibi}
\bibliography{references}

\onecolumngrid

\clearpage

\begin{center}
\textbf{\large End Matter}
\end{center}

\twocolumngrid

{\bf\textit{Coprecessions maximize corotating energy}}---We now prove that the corotating energy $\mathcal{H}_0[\vec{P}_\bp]$ is maximal if $\alpha_\bp>0$. Consider a small deviation from an exact coprecession satisfying \smash{$\vec{P}_\bp=\alpha_\bp \vec{H}_\bp$}. To first order, \smash{$\delta\vec{P}_\bp$} is orthogonal to $\vec{H}_\bp$ to conserve the length of \smash{$\vec{P}_\bp$}. Imposing this conservation to second order implies
\begin{equation}
    \vec{P}_\bp=\left(\alpha_\bp-\frac{|\delta\vec{P}_\bp|^2}{2\alpha_\bp |\vec{H}_\bp|^2}\right)\vec{H}_\bp+\delta\vec{P}_\bp.
\end{equation}
Recall that all quantities are understood in the frame of the unperturbed coprecession.

To first order, $\delta \mathcal{H}_0=\sum_\bp \vec{H}_\bp \cdot \delta\vec{P}_\bp$ vanishes as shown in the main text, whereas to second order
\begin{equation}\label{eq:change_energy}
    \delta\mathcal{H}_0=\frac{1}{2}\left(|\delta\vec{P}_0|^2-|\delta\vec{P}_1|^2\right)-\sum_\bp \frac{|\delta\vec{P}_\bp|^2}{2\alpha_\bp}.
\end{equation}
To make its sign manifest, note first the existence of the Goldstone mode $\delta\vec{P}_\bp^G=\delta \Phi\, \vec{B}\times \vec{P}_\bp$. As a pure phase rotation, this mode costs no energy. Its magnitude is $|\delta\vec{P}_\bp^G|=|\delta \Phi T_\bp|$, where $T_\bp$ is the signed component of $\vec P_\bp$ perpendicular to $\vec B$ within their common coprecession plane. The consistency condition $T_\bp=\alpha_\bp(T_0-vT_1)$ allows us to trade $\alpha_\bp$ for $T_\bp$ so that
\begin{equation}
    \delta \mathcal{H}_0=\sum_{\bp,\bp'}\frac{\delta\vec{P}_\bp\cdot \delta\vec{P}_{\bp'}-|\delta\vec{P}_\bp|^2 T_{\bp'}/T_\bp}{2}\,(1-vv').
\end{equation}
This last form can be rewritten as
\begin{equation}
    \delta \mathcal{H}_0=-\frac{1}{4}\sum_{\bp,\bp'}\left|\frac{\delta\vec{P}_\bp}{T_\bp}-\frac{\delta\vec{P}_{\bp'}}{T_{\bp'}}\right|^2(1-vv') T_\bp T_{\bp'}.
\end{equation}
This quadratic form is negative-semidefinite if all $T_\bp$ have the same sign. The fanning-out picture in Fig.~\ref{fig:scheme} makes this property geometrically evident: fanning out of an aligned configuration leaves all polarization vectors on the same side of the mass axis.

Define the angular moments $\alpha_n=\sum_\bp \alpha_\bp v^n$ and $T_n=\sum_\bp T_\bp v^n$. Taking the first moment of the consistency condition $T_\bp=\alpha_\bp(T_0-vT_1)$ gives $T_1=T_0\,\alpha_1/(1+\alpha_2)$ and therefore
\begin{equation}\label{eq:T-relation}
    T_\bp=\alpha_\bp\left(1-\frac{\alpha_1 v}{1+\alpha_2}\right)T_0.
\end{equation}
Summing this equation over $\bp$ then yields
\begin{equation}
    (\alpha_0-1)(\alpha_2+1)-\alpha_1^2=0.
\end{equation}
For $\alpha_\bp>0$, the Cauchy-Schwarz inequality $\alpha_1^2<\alpha_0\alpha_2$ gives
$\alpha_1^2<\alpha_2(1+\alpha_2)$ and finally
\begin{equation}
    \frac{\alpha_1^2}{(1+\alpha_2)^2}<\frac{\alpha_2}{1+\alpha_2}<1.
\end{equation}
Since $|v|\leq1$, Eq.~\eqref{eq:T-relation} then shows that all $T_\bp$ have the same sign. Hence $\delta \mathcal{H}_0$ is indeed negative-semidefinite and so $\mathcal{H}_0$ is maximal.

{\bf\textit{Microscopic proof of equation \eqref{eq:change_in_energy_finite}}}---We next prove 
$2\Delta \mathcal{H}_0=\Delta \Omega_{\rm c}\Delta  L_0-\Delta K_{\rm c}\Delta L_1$ from the microscopic equations. A neighboring coprecession satisfying the consistency equation obeys
\begin{eqnarray}\label{eq:Delta-EOM}
 \kern-2em \vec{P}_\bp+\Delta\vec{P}_\bp&=&(\alpha_\bp+\Delta \alpha_\bp)\Bigl[\vec{H}_\bp+\Delta\vec{P}_0-v\Delta\vec{P}_1
 \nonumber\\
 \kern-2em&&\kern6em{}-(\Delta \Omega_{\rm c}-v\Delta K_{\rm c})\vec{B}\Bigr].
\end{eqnarray}
Notice the distinction between the earlier generic perturbation $\delta\vec P_\bp$ about an exact coprecession and the present displacement \smash{$\Delta\vec P_\bp$} between nearby exact coprecessions. Linearizing Eq.~\eqref{eq:Delta-EOM} and multiplying by $\Delta\vec{P}_\bp$ yields
\begin{equation}
    \frac{|\Delta\vec{P}_\bp|^2}{\alpha_\bp}-\Delta\vec{P}_\bp\cdot(\Delta\vec{P}_0-v\Delta\vec{P}_1)=-(\Delta \Omega_{\rm c}-v\Delta K_{\rm c})\vec{B}\cdot \Delta\vec{P}_\bp.
\end{equation}
Summing over $\bp$ and specializing Eq.~\eqref{eq:change_energy} to 
\smash{$\delta\vec{P}_\bp=\Delta\vec{P}_\bp$}
immediately yields Eq.~\eqref{eq:change_in_energy_finite}.

{\bf\textit{Hydrodynamic evolution from the spin action}}---There is a yet more fundamental reason for the stationarity of $\mathcal{H}_0$. While $\mathcal{E}$ is not conserved, the equations of a system of interacting spins always descend from a spin action. For each mode, we write $P_\bp=|\vec P_\bp|$ and let $\varphi_\bp$ be its rotation phase around \smash{$\vec B$}. The spin action can then be expressed through this phase and the projection $P_{\parallel,\bp}$, without fixing any flavor-space coordinate system:
\begin{equation}
\begin{aligned}
\mathcal{S}[\vec P_\bp]={}&-\int dt\,dz\sum_\bp(P_\bp-P_{\parallel,\bp})\\
&\qquad\times(\partial_t+v\partial_z)\varphi_\bp
-\int dt\,dz\,\mathcal E[\vec P_\bp].
\end{aligned}
\end{equation}
Varying this action yields the equations of motion. The first term is the geometrical phase accumulated by each isospin as it evolves in space and time. For a coprecession, with fixed $P_{\parallel,\bp}$ and $\varphi_\bp=\Omega_{\rm c}t-K_{\rm c}z$, the Lagrangian density becomes $\mathcal{L}=-\mathcal E-\sum_\bp(P_\bp-P_{\parallel,\bp})(\Omega_{\rm c}-vK_{\rm c})$ or
\begin{equation}
    \mathcal{L}=-\mathcal{H}[\vec{P}_\bp;\Omega_{\rm c},K_{\rm c}]-\sum_\bp P_\bp(\Omega_{\rm c}-v K_{\rm c}).
\end{equation}
Since the last term is constant, the Lagrangian is the negative corotating energy up to a constant.

In the hydrodynamic limit, the exact action reduces to the effective action of the main text. 
Consider a slowly perturbed coprecession, with $\varphi_\bp=\Phi+\delta\varphi_\bp$ and $P_{\parallel,\bp}=P_{\parallel,\bp}^{(0)}+\delta P_{\parallel,\bp}$, where $\Phi$ is the collective phase. The longitudinal projections extremize the action
\begin{equation}\label{eq:variation}
    \frac{\delta \mathcal{E}}{\delta P_{\parallel,\bp}}= (\partial_t+v\partial_z)\Phi.
\end{equation}
If $\Phi=\Omega_{\rm c, 0}t-K_{\rm c, 0}z+\delta \Phi(t,z)$, perturbing Eq.~\eqref{eq:variation} gives
\begin{equation}
    \sum_{\bp'}\frac{\delta^2 \mathcal{E}}{\delta P_{\parallel,\bp}\,\delta P_{\parallel,\bp'}}\,\delta P_{\parallel,\bp'}=(\partial_t+v\partial_z)\delta \Phi.
\end{equation}
Defining $\mathcal{K}_{\bp,\bp'}=\delta^2\mathcal{E}/\delta P_{\parallel,\bp}\,\delta P_{\parallel,\bp'}$, its  inverse gives
\begin{equation}\label{eq:solution_delta_P}
    \delta P_{\parallel,\bp}=\sum_{\bp'}\mathcal{K}^{-1}_{\bp,\bp'}(\partial_t+v'\partial_z)\delta\Phi.
\end{equation}
By stationarity, the variation linear in \(\delta\Phi\) vanishes, while the quadratic term is 
\begin{equation}\label{eq:Seff-Appandix}
    \mathcal{S}_{\rm eff}=\frac{1}{2}\int\!dt dz \sum_{\bp,\bp'}(\partial_t\delta\Phi+v\partial_z\delta\Phi)\mathcal{K}^{-1}_{\bp,\bp'}(\partial_t\delta\Phi+v'\partial_z\delta\Phi).
\end{equation}
From here we can determine the coefficients $\Upsilon_{nm}$ for $n,m=0,1$ defined in the main text, and more generally the tensor $\Upsilon^{\mu\nu}=\sum_{\bp,\bp'} \mathcal{K}_{\bp,\bp'}^{-1} v^\mu v'^\nu$, where we introduce the four-component velocity $v^\mu=(1,\bv)$. So Eq.~\eqref{eq:Seff-Appandix} reproduces the effective action of the main text. Moreover, since the quadratic variation $\delta \mathcal{H}_0$ is negative-semidefinite, so is $\mathcal{K}$ and hence $\mathcal{K}^{-1}$ and $\Upsilon$. Hence the discriminant of Eq.~\eqref{eq:HydroDispersion} is generically negative, providing a remarkably elegant proof of instability. The form of $\Upsilon^{\mu\nu}$ also makes it clear that the instability vanishes for a single velocity mode: if all $v^\mu$ coincide, $\Upsilon^{\mu\nu}\propto v^\mu v^\nu$ has rank one.

{\bf\textit{Explicit construction of two-mode solution}}---The dispersion relation can be determined explicitly for an example of two modes $\vec{P}_{a}$ and $\vec{P}_{b}$, with results shown in Fig.~\ref{fig:scheme} for a family of solutions with different $\mu$. We here restore $\mu$ explicitly and fix units by $w_a=1$, with $w_b=0.4$, considering a spatial coprecession with $\Omega_{\rm c,0}=0$. Moreover, $|\vec{P}_{a}|=|\vec{P}_{b}|=1$, $v_a=1$, and $v_b=0.9$. Therefore, the total lepton-number flux $ L_1=v_aP_{\parallel,a}+v_bP_{\parallel,b}=1.6$ is held fixed as $\mu$ is varied.

For each choice of $\mu$, we determine the coprecession by solving the consistency condition \smash{$\vec{P}_\bp=\alpha_\bp \vec{H}_\bp$}. Using the signed transverse components $T_\bp$ and longitudinal projections $P_{\parallel,\bp}$, with $T_\bp^2+P_{\parallel,\bp}^2=P_\bp^2$, we obtain a coordinate-independent condition that eliminates $\alpha_\bp$:
\begin{equation}
\begin{aligned}
&T_\bp\biggl[w_\bp+K_{\rm c,0}v_\bp
+\mu\sum_{\bp'}P_{\parallel,\bp'}(1-vv')\biggr]\\
&\qquad=\mu P_{\parallel,\bp}\sum_{\bp'}T_{\bp'}(1-vv').
\end{aligned}
\end{equation}
Together with the fixed lepton-number flux $L_1$, these are three equations for $T_a$, $T_b$, and the coprecession wavenumber $K_{\rm c,0}$, with $P_{\parallel,a}$ and $P_{\parallel,b}$ determined by the fixed vector lengths and the chosen branches.

In the limit $\mu\gg w_\bp$, perfect alignment obtains, $\vec P_a/P_a\simeq\vec P_b/P_b$, corresponding to the synchronized regime. In the opposite limit, a continuous spectrum approaches a spectral split, with polarization vectors aligned or antialigned with the mass axis. For a discrete spectrum, this configuration usually cannot reproduce the fixed lepton-number flux. In our example, \smash{$\vec{P}_a$} aligns with the mass axis, while $\vec{P}_b$ remains inclined.

To determine the unstable modes of the coprecession, we introduce an oriented unit normal $\vec N$ to the coprecession plane and $\vec n_\bp=\vec P_\bp/P_\bp$. Without choosing any flavor-space coordinate axes, we decompose the length-preserving perturbation as $\delta\vec P_\bp=\xi_\bp(\vec N\times\vec n_\bp)+\eta_\bp\vec N$. Thus $\xi_\bp$ and $\eta_\bp$ are its in-plane and out-of-plane amplitudes, respectively. Then the equations of motion are
\begin{equation}
\begin{aligned}
(\partial_t+v\partial_z)\xi_\bp
&=\mu P_\bp(\eta_0-v\eta_1)-H_\bp\eta_\bp,\\
(\partial_t+v\partial_z)\eta_\bp
&=H_\bp\xi_\bp-\mu(\vec\xi_0-v\vec\xi_1)\cdot\vec P_\bp,
\end{aligned}
\end{equation}
where $\eta_n=\sum_\bp\eta_\bp v^n$ and $\vec\xi_n=\sum_\bp\xi_\bp\vec n_\bp v^n$. For two modes, one obtains a four-by-four system that can be diagonalized numerically, yielding the growth rates shown in the bottom panel of Fig.~\ref{fig:scheme}.

{\bf\textit{Generalized flavor susceptibility}}---We now provide an explicit expression for the generalized flavor susceptibility. The kinetic perturbation \smash{$\delta \vec{P}_\bp$} in Eq.~\eqref{eq:explicit_solution} responds to the collective field \smash{$(\delta \vec{P}_0,\delta \vec{P}_1)$}. Beyond axial symmetry, this field is promoted to a four-vector through \smash{$\delta \vec{P}^\mu=\sum_\bp \delta \vec{P}_\bp v^\mu$}, carrying both a Lorentz and an isospin index, $v^\mu=(1,\bv)$ with $\bv=\bp/|\bp|$, and the perturbation has wavevector $K^\mu=(\omega,\bk)$. 
Flavor-isospin indices have a Euclidean metric, so their upper or lower positions are equivalent. The flavor susceptibility is thus  a four-index tensor such that $\delta P^{\mu i}=\chi^{\mu i}_{\nu j} \delta P^{\nu j}$. From  Eq.~\eqref{eq:explicit_solution} we infer
\begin{equation}\label{eq:suceptibility}
   \chi^{\mu i}_{\nu j}=\sum_\bp \frac{\alpha_\bp\bigl[H_{\bp}^i H_{\bp}^j-|\vec{H}_\bp|^2 \delta_{ij}+i(K\cdot v) \varepsilon_{ijk} H_{\bp}^k\bigr]}{(K\cdot v)^2-|\vec{H}_\bp|^2}\,v^\mu v_\nu ,
\end{equation}
where $K\cdot v=\omega-\bv\cdot\bk$. The dispersion relation then has a familiar form in its Lorentz-index structure, but includes also flavor isospin indices $\mathrm{det}\bigl[\delta^\mu_\nu \delta^i_j-\chi^{\mu i}_{\nu j}\bigr]=0$.

The resonant denominators in Eq.~\eqref {eq:suceptibility} directly reveal the flavor-cyclotron resonances at \smash{$K\cdot v=\pm|\vec H_\bp|$}, corresponding to resonant wave-particle interaction. The Levi-Civita tensor $\varepsilon_{ijk}$ introduces the antisymmetric part of the response that is allowed by the axial symmetry (in isospin space) around the external field \smash{$\vec{H}_{\bp}$}. 

We notice again the strong similarity with a plasma in a magnetic field. The susceptibility can generically be written in terms of three structure functions and the external field, here expressed through the intensity $\bH$,
\begin{equation}
    \chi_{ij}=\chi_S(H^2 \delta_{ij}-H_i H_j)+\chi_P H_i H_j +i \chi_A \varepsilon_{ijk}H_k.
\end{equation}
The flavor-magnetic susceptibility, by contrast, contains only the first and third structure function, at least for our studied case of flavor coprecession. The second is absent by virtue of conservation of the length of $\vec{P}_\bp$, preventing perturbations parallel to the local field $\vec{H}_\bp$. In a magnetized plasma, particles can instead accelerate along the magnetic field through parallel electric fluctuations.

\onecolumngrid

\clearpage

\onecolumngrid
\appendix

\setcounter{equation}{0}
\setcounter{figure}{0}
\setcounter{table}{0}
\setcounter{page}{1}
\makeatletter
\renewcommand{\theequation}{S\arabic{equation}}
\renewcommand{\thefigure}{S\arabic{figure}}
\renewcommand{\thepage}{S\arabic{page}}

\begin{center}
\textbf{\large Supplemental Material for the Letter\\[0.5ex]
{\em Phase-Twist Instability in a Coprecessing Neutrino Plasma}}
\end{center}

\bigskip

We review (A)~the dispersion relation of magnetized plasmas and explain (B)~the firehose instability by magneto-hydrodynamical reasoning alone. (C)~We further compare our results for the phase-twist instability with a previous two-beam example of a flavor-nondiagonal setup. (D)~We prove the general stability of aligned polarization vectors through a new flavor-helicity argument, which also recovers the usual crossing condition for flavor-diagonal plasmas.
(E)~We study the instability condition for modes that break local axial symmetry.

\bigskip

\twocolumngrid

\section{A. Collisionless magnetized plasmas for pedestrians}

We review the dispersion relation of a nonrelativistic collisionless magnetized plasma consisting of protons and electrons. For a particle species of charge $e$, mass $m$, and phase-space distribution $f_\bp(t,\br)$, the kinetic equation is
\begin{equation}
    \partial_tf_\bp+\bv\cdot \partial_\br f_\bp+e(\bE+\bv\times \bB)\cdot\partial_\bp f_\bp=0,
\end{equation}
where $\bE$ and $\bB$ are the electric and magnetic fields, and $\bv=\bp/m$. An equation of this form applies to each charged species. Plasma neutrality requires that the unperturbed proton and electron densities are equal.

For a homogeneous and stationary background with magnetic field $\bB_0$, the unperturbed transport equation becomes
\begin{equation}
    (\bv\times \bB_0)\cdot \partial_\bp f_{0,\bp}=0.
\end{equation}
This equation implies that $f_{0,\bp}$ must be axisymmetric around $\bB_0$, but can be any function of zenith angle.

A small electromagnetic perturbation $\be$ and $\bb$ with plane-wave variation $e^{-i\omega t+i\bk\cdot \br}$ leads to a perturbation $\delta f_\bp$ obeying
\begin{equation}\label{eq:f0-kinetic}
    i(\bk\cdot\bv-\omega)\delta f_\bp+e(\bv\times \bB_0)\cdot \partial_\bp\delta f_\bp=-e(\be+\bv\times\bb)\cdot\partial_\bp f_{0,\bp}.
\end{equation}
Notice that Maxwell's equations imply $\bb=\bk\times \be/\omega$.

The perturbation $\delta f_\bp$ need not be axisymmetric around $\bB_0$. In terms of the azimuthal angle $\varphi$ and the cyclotron frequency $\omega_B=eB_0/m$, Eq.~\eqref{eq:f0-kinetic} becomes
\begin{equation}
    i(\bk\cdot \bv-\omega)\delta f_\bp-\omega_B \partial_\varphi\delta f_\bp=-e\left[\be+\frac{\bv\times(\bk\times\be)}{\omega}\right]\cdot\partial_\bp f_{0,\bp}.
\end{equation}
We restrict our study to perturbations with $\bk\parallel \bB_0$; the more general solution is found, e.g.\ in Ref.~\cite{Bernstein:1958zz}. We then expand in azimuthal harmonics $\delta f_\bp=\sum_{s=-\infty}^{+\infty} g_s e^{i s \varphi}$. 

It is convenient to separate the responses to longitudinal and transverse perturbations. With $\bB_0$ and $\bk$ defining the $z$ axis, the longitudinal field is $e_z$, whereas the transverse field is expressed in terms of the circularly polarized variables $e_\pm=e_x\pm i e_y$. The solution then takes the form
\begin{equation}\label{eq:perturbed_distribution}
    \delta f_\bp=-ie\!\left[\frac{e_z\partial_{p_z}f_{0,\bp}}{\omega-\bk\cdot \bv}+\mathcal{D}_p f_{0,\bp}\sum_{\sigma=\pm}\frac{e_\sigma e^{-i\sigma \varphi}}{2(\omega-\bk\cdot\bv-\sigma \omega_B)}\right]
\end{equation}
with the derivative
\begin{equation}
    \mathcal{D}_p=\left(1-\frac{\bk\cdot \bv}{\omega}\right)\partial_{p_\perp}+\frac{k v_\perp}{\omega}\partial_{p_z}.
\end{equation}
Here $p_\perp$ and $v_\perp$ are the magnitudes of the momentum and velocity transverse to the field.

Longitudinal perturbations thus excite conventional Cherenkov resonances associated with plasmon emission and absorption, while transverse perturbations excite cyclotron resonances. Since longitudinal modes propagating along $\bB_0$ behave identically to the unmagnetized plasma that we have reviewed in Ref.~\cite{Fiorillo:2024bzm}, we here focus on the transverse, circularly polarized modes. 

To obtain a closed dispersion relation, we relate the perturbed current density to the electric field that generates it. This is analogous to the self-consistency step in neutrino plasmas, where we relate $\delta\vec{P}_\mu$ to the individual \smash{$\delta\vec{P}_\bp$} through \smash{$\delta\vec{P}_\mu=\sum_\bp \delta\vec{P}_\bp v^\mu$}. In the present case, the relevant relation is Maxwell's fourth equation
\begin{equation}
    \boldsymbol{\nabla}\times \bB=4\pi \mathbf{j}+\partial_t \bE.
\end{equation}
For our plane, monochromatic, circular modes, this equation becomes
\begin{equation}
    e_+\left(1-\frac{k^2}{\omega^2}\right)+\frac{4\pi i}{\omega}j_+=0,
\end{equation}
where $j_+=j_x+ij_y$. The latter follows from the perturbed distribution in Eq.~\eqref{eq:perturbed_distribution} by multiplication with $v_x+i v_y=v_\perp e^{i\varphi}$ and integration over phase space. We thus find the dispersion relation
\begin{equation}
    1-\frac{k^2}{\omega^2}+\sum_{\alpha=p,e}\frac{2\pi e_\alpha^2}{\omega}\int \frac{d^3\bp}{(2\pi)^3}\frac{ v_\perp\mathcal{D}_p f_{0\alpha,\bp}}{\omega-kv_z-\omega_{B,\alpha}}=0,
\end{equation}
where the sum is over protons and electrons.

One could now examine this dispersion relation for a given distribution $f_{0\alpha,\bp}$. If it is sufficiently anisotropic, one finds the firehose instability at long wavelengths~\cite{Parker:1958zzb}. It occurs when the pressure along the magnetic field sufficiently exceeds the transverse pressure. The anisotropy can then overcome the restoring magnetic tension, making transverse perturbations unstable and leading to a buckling of the field lines---hence the name. Since this is essentially a long-wavelength effect, its underlying physics becomes far more transparent in a hydrodynamical rather than a kinetic description. At shorter scales, however, kinetic effects become important and eventually provide stabilization.

\section{B. Firehose instability from magnetohydrodynamics}

Intriguingly, the firehose instability can be found by a route that mirrors directly the hydrodynamical approach of the main text for the flavor version of this instability. The key idea is that the kinetic treatment can be replaced by a hydrodynamical one, based on macroscopic conservation laws, when $\omega$ and $k v_{\rm th}$ are much smaller than the cyclotron frequencies $\omega_{B\alpha}$ of all species. Here, $v_{\rm th}$ is a typical particle velocity and $(\omega,k)$ is the studied perturbation.

It is important to note that we are not assuming the presence of fast collisions that would restore thermal equilibrium. Rather, the hydrodynamical time scale is long compared with the cyclotron period $\omega_{B\alpha}^{-1}$. As a result, within each fluid element the distribution need not be isotropic or thermal. As we have seen in the previous section, it need only be axisymmetric around the magnetic-field direction.

We shall take only the simplest setup, consisting of an incompressible magnetized fluid, composed of electrons and protons, described by a single bulk velocity $\bU$, while their relative drift produces the current $\bj$. Newton's law therefore reads
\begin{equation}\label{eq:newton}
    \rho\bigl[\partial_t \bU + (\bU\cdot \boldsymbol{\nabla})\bU\bigr]=-\boldsymbol{\nabla}\cdot \boldsymbol{P}+\mathbf{j}\times \bB,
\end{equation}
where $\rho$ is the mass density and $\boldsymbol{P}$ is the stress tensor. For an unmagnetized fluid, $\boldsymbol{P}$ is usually taken to be isotropic and simply corresponds to the pressure. 

However, in a magnetized collisionless fluid, the distribution need not be isotropic. We therefore assume a separate pressure along and transverse to the magnetic field direction
\begin{equation}
    P_{ij}=P_\parallel b_i b_j + P_\perp (\delta_{ij}-b_i b_j),
\end{equation}
where $\bb=\bB/|\bB|$ is the unit vector along the field. 

We finally assume the plasma is ideal, with vanishing resistivity. The direct implication is that $\bE=-\bU\times\bB$, which implies
\begin{equation}
    \partial_t \bB=\boldsymbol{\nabla}\times(\bU\times \bB).
\end{equation}
The induction equation, together with Newton’s law and the anisotropic stress tensor, completes our set of hydrodynamical assumptions.

We now have all the ingredients to predict the firehose instability. We consider a uniform background field $\bB_0$, perturbed as
$\bB=\bB_0+\delta \bB$. The background fluid is stationary, so $\bU$ is a perturbation at the same order as $\delta \bB$. The two are related by the third Maxwell equation
\begin{equation}
    \partial_t \delta \bB=\boldsymbol{\nabla}\times(\bU\times \bB_0).
\end{equation}
For a plane monochromatic perturbation evolving with the harmonic factor $e^{-i\omega t+i\bk\cdot \br}$, we find
\begin{equation}
    \delta\bB=-\frac{\bk\times(\bU\times\bB_0)}{\omega}.
\end{equation}
The medium current in response follows from Maxwell's fourth equation, in which we may neglect $\partial_t \bE$ in the nonrelativistic limit. Therefore we find
\begin{equation}
    \mathbf{j}=\frac{\boldsymbol{\nabla}\times \bB}{4\pi}=\frac{i\bigl[\bk\times\bB_0(\bU\cdot\bk)-\bk\times\bU (\bk\cdot\bB_0)\bigr]}{4\pi\omega}.
\end{equation}
Thus both $\delta\bB$ and $\bj$ are determined by the fluid-velocity perturbation $\bU$.

To simplify the discussion, we revert to the assumption that $\bk\parallel \bB_0$ and $\bU\perp \bB_0$, as in the kinetic case, so that the current reduces to
\begin{equation}
    \bj=i\bU\times \bk\frac{k B_0}{4\pi\omega}.
\end{equation}
Finally, the term $\boldsymbol{\nabla}\cdot \boldsymbol{P}$ is also perturbed because of the buckling of the field lines: the unit vector $\mathbf{b}$ tilts under the perturbation. By explicitly perturbing it, we find
\begin{equation}
    \delta \mathbf{b}=-\frac{k \bU}{\omega}.
\end{equation}
Therefore, the perturbed divergence of the stress tensor is
\begin{equation}
    \boldsymbol{\nabla}\cdot \boldsymbol{P}=-\frac{ik^2(P_\parallel-P_\perp)}{\omega}\,\bU.
\end{equation}
Upon replacing these perturbations in Eq.~\eqref{eq:newton}, we are finally led to the hydrodynamical dispersion relation,
\begin{equation}
    \omega^2=\frac{k^2}{\rho}\left(\frac{B_0^2}{4\pi}+P_\perp-P_\parallel\right)=u^2 k^2
\end{equation}
where $\omega=u k$ defines the phase velocity $u$.

For real phase velocity $u$, corresponding to $u^2>0$, this dispersion relation describes stable waves. They are simply Alfv\'en waves, as the limit $P_\perp=P_\parallel$ reveals. However, for $P_\parallel>P_\perp + B_0^2/4\pi$, $u$ becomes imaginary. The dispersion relation then takes exactly the same form as for the flavor-firehose hydrodynamical instability of the main text, with the growth rate increasing linearly with $k$. As in the flavor case, the linear increase does not continue forever because the hydrodynamical approximation breaks down when $\omega$ or $k v_{\rm th}$ become comparable with the cyclotron frequencies.

Overall, we thus find nearly identical pictures for the firehose and phase-twist instabilities.
In the magnetic case, a sufficiently large pressure anisotropy $P_\parallel-P_\perp$ destabilizes the field lines through slow buckling, just as the slow modulation of the coprecession plane becomes unstable when the polarization vectors are misaligned. The instability emerges from the hydrodynamical modes: Alfv\'en waves in the magnetized plasma and flavor-precession waves in the flavor-nondiagonal plasma. At sufficiently large wavenumbers, kinetic corrections set in, causing the growth rate to turn over as a function of $k$ and eventually vanish in both cases.

\section{C. Relation to previous flavor-nondiagonal analyses}

The idea that nonlinear flavor relaxation could produce a slowly varying flavor-nondiagonal background led Johns and Kost~\cite{Johns:2025yxa} to propose extending the concept of flavor waves to disturbances on generic flavor-nondiagonal backgrounds. They formally linearized around such a generic background. More recently, Kost, Johns and Duan~\cite{Kost:2026ckc} studied a fast-flavor toy example of two colliding beams with initially nonaligned polarization vectors. For a fixed wavenumber, they identified an instability that depends on the chosen lengths and relative orientation of the two polarization vectors. 

With the benefit of hindsight from our analysis, one can recognize that this example is surprisingly analogous to our own two-mode system. More generally, coprecessions also exist in the fast-flavor limit, but unlike in the slow-flavor case, they do not arise naturally as an adiabatically connected family when $\mu$ is varied. The two-mode system is special, however, in that its homogeneous motion is always a coprecession. Its rotation axis is fixed by initial conditions rather than an external vector and therefore depends on the initial polarization vectors.

To make the connection explicit, consider the two-beam fast-flavor system studied in Ref.~\cite{Kost:2026ckc}, consisting of left- and right-moving modes with polarization vectors $\vec P_L$ and $\vec P_R$. Their equations of motion are
\begin{eqnarray}\label{eq:evolution_two_beam}
    (\partial_t-\partial_z)\vec{P}_L&=&2 \vec{P}_R\times \vec{P}_L,\\
    (\partial_t+\partial_z)\vec{P}_R&=&2 \vec{P}_L\times \vec{P}_R,
\end{eqnarray}
where the interaction strength $\mu$ was absorbed in the units of space and time. Reference~\cite{Kost:2026ckc} then considered the family of backgrounds
\begin{equation}
    \vec{P}_L=(P_T,0,0.1),\; \vec{P}_R=(P_T,0,0.5)
\end{equation}
and diagonalized the linearized system numerically at the fixed wavenumber $k/\mu\simeq0.021$ and different $P_T$.

For a homogeneous fast flavor system, the sum of all polarization vectors plays the role of a conserved angular momentum that defines an axis in flavor space. For two modes, this is $\vec{S}=\vec{P}_L+\vec{P}_R$ and both of them obey \smash{$\partial_t\vec{P}_{L,R}=2 \vec{S}\times \vec{P}_{L,R}$}. Thus both vectors precess with the same angular velocity around $\vec S$, which lies in the plane they define, so the motion is a coprecession.  A larger number of modes does not generically coprecess. 

The two-mode system thus has a fixed point in the corotating frame. One can then linearize the equations to find the normal modes. Introducing small disturbances $\delta\vec{P}_{L,R}(t,z)$, they obey in the corotating frame
\begin{eqnarray}
    (\partial_t-\partial_z)\delta \vec{P}_L&=&2(\delta \vec{P}_R+\delta \vec{P}_L)\times \vec{P}_L,\\
    (\partial_t+\partial_z)\delta \vec{P}_R&=&2(\delta \vec{P}_R+\delta\vec{P}_L)\times \vec{P}_R.
\end{eqnarray}
With $\vec{S}=\vec{P}_L+\vec{P}_R$ and $\vec{D}=\vec{P}_L-\vec{P}_R$ and $\delta \vec{S}$ and $\delta \vec{D}$ their perturbations, and assuming plane-wave solutions, $e^{-i\omega t+ikz}$, we find
\begin{eqnarray}
    \omega \delta \vec{S}+k\delta\vec{D}&=&-2i \vec{S}\times \delta \vec{S},\\
    \omega \delta \vec{D}+k\delta\vec{S}&=&-2i\vec{D}\times \delta \vec{S}.
\end{eqnarray}
Eliminating $\delta \vec{D}$ gives
\begin{equation}
    (\omega^2-k^2)\delta \vec{S}=-2i(\omega \vec{S}-k\vec{D})\times \delta \vec{S}.
\end{equation}
Taking another cross product with $\omega\vec{S}-k\vec{D}$, we finally find the dispersion relation
\begin{equation}
    (\omega^2-k^2)^2=4[\vec{S}^2\omega^2-2\vec{S}\cdot \vec{D} \omega k+\vec{D}^2 k^2].
\end{equation}
One can solve this quartic equation for any choice of the two-beam background model $\vec{P}_{L,R}$.
For $k\to 0$, one finds a stable gapped mode with \smash{$\omega=2|\vec{S}|$}, which actually remains stable for any $k$.

Instead, we focus on the gapless unstable mode with $\omega=u_\pm k$ in the hydrodynamic limit. The left-hand side is of higher order in $k$ and may be neglected so that
\begin{equation}
    u_\pm=\frac{\vec{S}\cdot \vec{D}\pm \sqrt{(\vec{S}\cdot\vec{D})^2-S^2 D^2}}{S^2}.
\end{equation}
The argument of the square root is negative whenever the two polarization vectors are nonaligned, in agreement with our instability theorem. Our analytic growth rates agree with the numerical values found in Ref.~\cite{Kost:2026ckc}.

By going to the corotating frame, we have effectively absorbed its precession frequency into the real part of the mode frequency. The imaginary part remains the same, so in this special case, solving the equation in the corotating frame was not crucial for identifying the growth rate. 

This discussion clarifies the nature of the analysis in Ref.~\cite{Kost:2026ckc}. They evaluate the flavor-wave Hamiltonian directly for the prescribed background polarization vectors, which, however, for $P_T\neq0$ are not stationary in the used frame. Thus, its eigenvalues are not the normal-mode ones of the evolving background. Instead, they are the eigenvalues of an instantaneous snapshot of that background, and would be physically meaningful only if the background were quasi-static. Indeed, Ref.~\cite{Kost:2026ckc} explicitly notes this limitation. 

However, for the particular case of a fast two-beam system, fortuitously the growth rate turns out correct. The reason is that \textit{any} choice of the homogeneous background results in a coprecession~\cite{Fiorillo:2026jgw}, as we have seen. Therefore, in the corotating frame it is indeed static, and the transformation to this frame changes only the real part of the mode frequency. The reported growth rate of Ref.~\cite{Kost:2026ckc} therefore agrees with our analytic result. What was not discussed in Ref.~\cite{Kost:2026ckc} is the coprecession nature of the background---without it, the diagonalization of a generically nondiagonal background does not provide the correct growth rate of the system---and its connection to the general phase-twist instability that we have discovered here.

\onecolumngrid

\section{D. Instability criterion from flavor-helicity states}

Perfectly aligned polarization vectors with $\alpha_\bp>0$ are stable, i.e., when they all point in the same direction. This general result was proven in the main text by appealing to lepton number conservation. This argument is analogous to the necessity criterion~\cite{Dasgupta:2021gfs}, closely related to lepton-number conservation~\cite{Johns:2024esf, Fiorillo:2024bzm}, of a crossing for polarization vectors aligned with the $z$ direction. 

It is intriguing, however, that the dispersion relation contains this information directly. To this end, we introduce helical flavor states instead of Cartesian polarization vectors. In this way, one can show that the instability requires a nonvanishing coupling between states of opposite flavor helicity.

We start from Eq.~\eqref{eq:perturbation_evolution} for the perturbations around a generic coprecessing background. We introduce the complex variables $\psi_{\pm,\bp}=\xi_\bp\pm i\eta_\bp$, where we recall (see End Matter) that $\xi_\bp$ and $\eta_\bp$ are perturbations in the plane orthogonal to the unperturbed polarization vector $\vec{P}_\bp$, with $\eta_\bp$ normal to the coprecession plane. Assuming a space-time variation for the perturbation of $\psi_{\pm,\bp}\propto e^{-i\omega t +i kz}$, we find a remarkably symmetric structure,
\begin{eqnarray}\label{eq:helicityEOM1}
    (\omega-kv)\psi_{+,\bp}&=&-H_\bp \psi_{+,\bp}+\sum_{\bp'}(1-vv')\cos^2\left(\frac{\theta_{\bp\bp'}}{2}\right)P_\bp \psi_{+,\bp'}-\sum_{\bp'}(1-vv')\sin^2\left(\frac{\theta_{\bp\bp'}}{2}\right)P_\bp \psi_{-,\bp'},
    \\ \label{eq:helicityEOM2}
    (\omega-kv)\psi_{-,\bp}&=&+H_\bp \psi_{-,\bp}-\sum_{\bp'}(1-vv')\cos^2\left(\frac{\theta_{\bp\bp'}}{2}\right)P_\bp \psi_{-,\bp'}+\sum_{\bp'}(1-vv')\sin^2\left(\frac{\theta_{\bp\bp'}}{2}\right)P_\bp \psi_{+,\bp'},
\end{eqnarray}
where $H_\bp$ and $P_\bp$ are the lengths of $\vec{H}_\bp$ and $\vec{P}_\bp$, respectively. The angle in flavor space between two polarization vectors $\vec{P}_\bp$ and $\vec{P}_{\bp'}$ is $\theta_{\bp\bp'}$.

The perturbations described by $\psi_{\pm,\bp}$ are waves with opposite circular polarization, i.e., the field precesses oppositely in the $\xi$--$\eta$--plane. A crucial observation is that the couplings between opposite flavor-helicity states, the last terms in Eqs.~\eqref{eq:helicityEOM1} and~\eqref{eq:helicityEOM2}, vanish when all polarization vectors are aligned, corresponding to the limit $\theta_{\bp\bp'}\to 0$. The two flavor-helicity classes then evolve independently. 

It is then immediate that, in this aligned configuration, no instability can develop. Since $P_\bp>0$, we can express \smash{$\psi_{\pm,\bp}=\sqrt{P_\bp}\tilde{\psi}_{\pm,\bp}$} with an equation of motion
\begin{equation}
    (\omega-kv+H_\bp)\tilde{\psi}_{+,\bp}=\sum_{\bp'}(1-vv')\cos^2\left(\frac{\theta_{\bp\bp'}}{2}\right)\sqrt{P_\bp P_{\bp'}}\tilde{\psi}_{+,\bp'}.
\end{equation}
The matrix on the right-hand side is Hermitian, its eigenvalues are therefore real, and thus leaving no room for an instability. An instability can only arise through the $\sin^2(\theta_{\bp\bp'}/2)$ term which couples the two polarizations. 

This general argument also implies the well-known necessary criterion of crossing~\cite{Dasgupta:2021gfs} in flavor-diagonal plasmas. A crossing means that for some polarization vectors, $\theta_{\bp\bp'}=\pi$ applies, leading to a nonvanishing coupling between the left-handed and right-handed perturbations of oppositely oriented polarization vectors.

\section{E. Transverse-symmetry-breaking modes.}

Our discussion has centered on modes depending on time and on the longitudinal coordinate of the spectral split. We now briefly discuss the generalization of this theory to modes that break the transverse symmetry, with $\bk$ not aligned with the direction of the coprecession vector $\mathbf{K}_{\rm c}=(0,0,K_{\rm c})$. We maintain the assumption that the background distribution is azimuthally symmetric around this axis, so $P_\bp$ depends on energy $|\bp|$ and on $v$, but not on the azimuthal angle $\phi$ around the axis $z$.

The properties of the phase-twist instability in terms of $\bk=(k_T,0,k_L)$ can be deduced from the same dimensional arguments used in the main text. The hydrodynamic dispersion relation is, to a first approximation, $\omega\simeq u_R k_L$. We assume a small angular width $\Delta \theta$ for the polarization vectors, as in the main text, and a small angular spread for the neutrino directions $\Delta \vartheta$. As a result, we have $\Delta v\sim \Delta \vartheta^2$. The generalization of Eq.~\eqref{eq:dispersion_relation_hydro_expansion} must include all corrections depending on $k_L\Delta v\sim k_L \Delta \vartheta^2$ and on $k_T\Delta \vartheta$; the reason is that in the dispersion relation $k_L$ appears multiplied by $v$ and $k_T$ by $\sqrt{1-v^2}\cos \phi\sim \vartheta$. Therefore, we have
\begin{equation}
    (\omega-u_R k_L)^2=-\Delta \theta^2(a k_L^2\Delta \vartheta^4+bk_T^2\Delta \vartheta^2)+\frac{\alpha k_T^4 \Delta \vartheta^4+\beta k_L^4\Delta\vartheta^8+\sigma k_T^2 k_L^2 \Delta \vartheta^6}{H^2}+...
\end{equation}
Here $H$ is the typical magnitude of $H_\bp$, which we have extracted so that all the coefficients $a$, $b$, $\alpha$, $\beta$, $\gamma$ are dimensionless and expected of order unity.

\begin{figure}
    \centering
    \includegraphics[width=0.49\columnwidth]{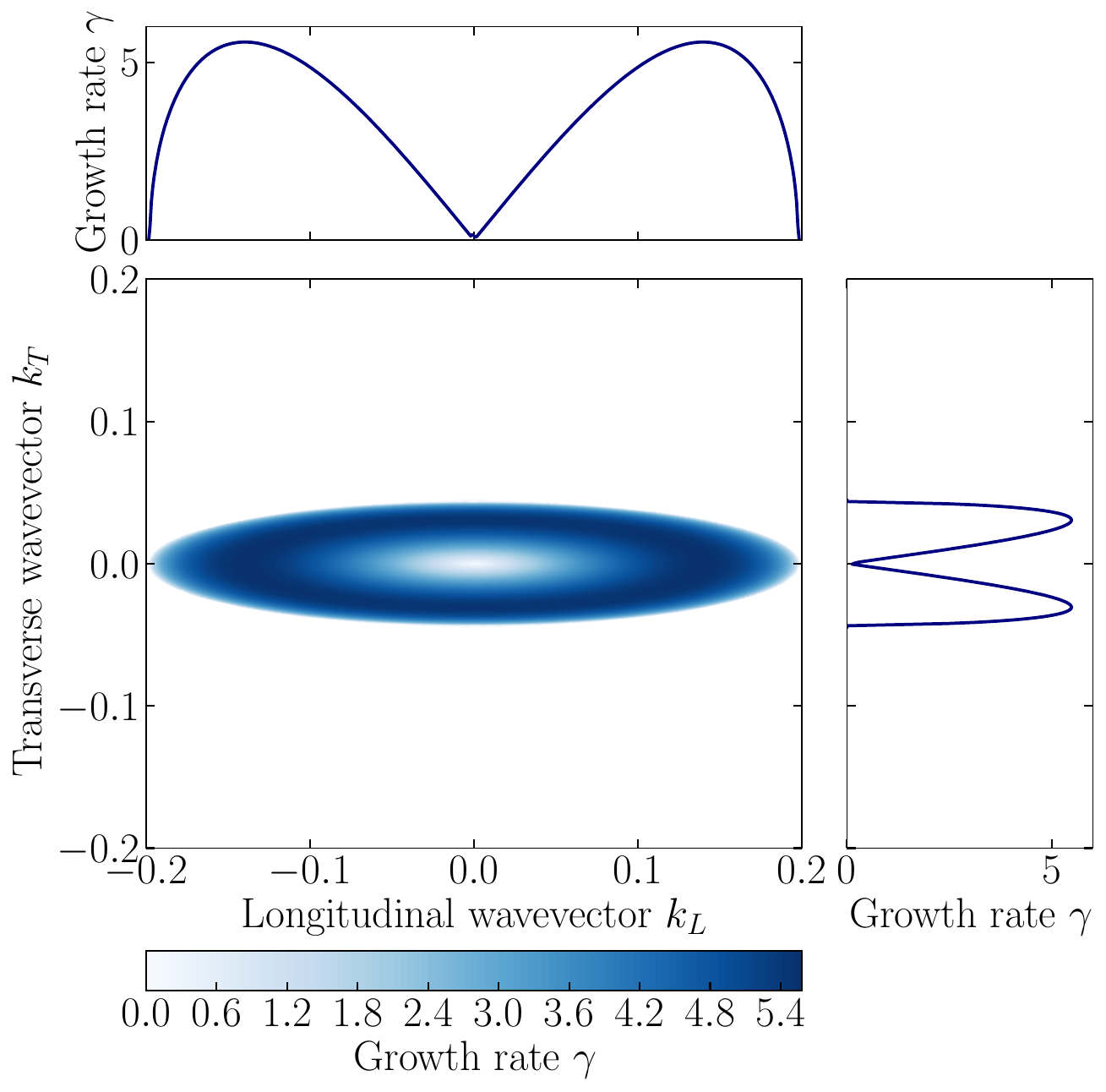}
    \caption{Growth rate as a function of the transverse and longitudinal components of the wavevector, for the second benchmark case ($\mu=15$) from Fig.~\ref{fig:scheme} in the main text.}
    \label{fig:growth_full}
    \vskip-6pt
\end{figure}

From here we see that, for small $\Delta \vartheta$, the growth rate grows much faster for transverse-breaking modes with $k_L=0$ than for longitudinal modes with $k_T=0$, since $k_T$ is multiplied by a lesser power of $\Delta \vartheta$. On the other hand, the instability disappears at a smaller $k_T$; the maximum value of the growth rate remains of the order of $\gamma_{\rm max}\sim H \Delta \theta^2$, and is achieved at $k_L\sim H \Delta \theta/\Delta\vartheta^2$ and $k_T\sim H \Delta \theta/\Delta \vartheta$.

We can verify explicitly these scalings, deduced theoretically, from an explicit diagonalization. The flavor susceptibility in Eq.~\eqref{eq:suceptibility} can be integrated exactly over the azimuthal angle. If we replace the sum over $\bp$ by 
\begin{equation}
    \sum_{E,v}=\int \frac{E^2 dE dv}{(2\pi)^2},
\end{equation}
the integrals over $\phi$ have all the form
\begin{equation}
    J_n=\int_0^{2\pi}\frac{d\phi}{2\pi}\frac{\cos^n\phi}{a-b \cos \phi},
\end{equation}
where $a=\omega-k_L v -\sigma |\vec{H}_\bp|$ with $\sigma=\pm 1$ and $b = \sqrt{1-v^2} k_T$; the first few integrals are
\begin{equation}
    J_0=\frac{1}{\sqrt{a^2-b^2}},\quad J_1=\frac{aJ_0-1}{b},\quad J_2=\frac{a J_1}{b}.
\end{equation}
Therefore, we can construct the matrix
\begin{equation}
    \mathcal{M}^\mu_\nu(a,b)=\int_0^{2\pi}\frac{d\phi}{2\pi} \frac{v^\mu v_\nu}{a-b\cos \phi}=\begin{pmatrix}
        J_0 & - \sqrt{1-v^2}J_1 & 0 & - v J_0 \\
        \sqrt{1-v^2}J_1 & (v^2-1) J_2 & 0 & - \sqrt{1-v^2} v J_1 \\
        0 & 0 & (v^2-1)(J_0-J_2) & 0\\
        v J_0 & - \sqrt{1-v^2} v J_1 & 0 & - v^2 J_0
    \end{pmatrix}.
\end{equation}
After separating the susceptibility in Eq.~\eqref{eq:suceptibility} in its $\sigma=\pm 1$ components to obtain
\begin{equation}
    \chi^{\mu i}_{\nu j}=\sum_{E,v}\frac{\alpha_\bp}{2}\sum_{\sigma=\pm 1}\left[\frac{\sigma (H_\bp^i H_\bp^j-|\vec{H}_\bp|^2 \delta^{ij})}{|\vec{H}_\bp|^2}+i\varepsilon^{ijk} H_\bp^k\right]\mathcal{M}^\mu_\nu(\omega-k_L v -\sigma |\vec{H}_\bp|,\sqrt{1-v^2}k_T).
\end{equation}
We solve this dispersion relation numerically for the benchmark case shown in Fig.~\ref{fig:scheme} in the main text with $\mu=15$, and show the resulting growth rate for arbitrary $(k_T,k_L)$ in Fig.~\ref{fig:growth_full}.

The results completely confirm our theoretical expectations. The growth rate is non-vanishing roughly in an ellipse in $\bk$-space, which is considerably more elongated along the longitudinal perturbations $k_L$ than the transverse ones $k_T$. This is in line with our expectations, that the instability survives to smaller values of $k_T$ than $k_L$ by a factor $\Delta \vartheta\sim 0.1$ in our benchmark case. Nevertheless, the maximum growth rate is essentially the same along all directions, matching again our prediction from the hydrodynamical expansion.

\onecolumngrid



\end{document}